\documentclass[aps,pra,twocolumn,groupedaddress,showpacs,floatfix]{revtex4-1}
\usepackage{graphicx}
\usepackage{braket}
\usepackage{bm}
\usepackage{color}

\begin{document}
\title{Excitation Suppression Due to Interactions Between Microwave-Dressed Rydberg Atoms}
\author{E. Brekke}
\affiliation{Department of Physics, St. Norbert College, De Pere, WI 54115}
\author{J. O. Day}
\affiliation{University of Wisconsin-Madison}
\altaffiliation{Current address: U.S. House of Representatives}
\author{T. G. Walker}
\affiliation{Department of Physics, University of Wisconsin-Madison, Madison, WI 53706}
\date{\today}
\begin{abstract}
Atom-atom interactions within a small volume were investigated through the excitation of ultracold Rb atoms. The application of microwaves enhances these interactions, causing the suppression of Rydberg state excitation. The suppression of Rydberg atom excitation was both qualitatively observed and quantitatively analyzed using a universal scaling law, giving a measure of the atom-atom interaction strength in agreement with theoretical prediction.
\end{abstract}
\maketitle

\section{Introduction}
Strong dipole-dipole interactions between atoms in Rydberg states allow conditional quantum manipulations of atom pairs at micron-scale interatomic separations \cite{Saffman2010a}.  The blockade effect \cite{Jaksch2000} uses these interactions to generate entanglement, as recently demonstrated experimentally \cite{Wilk2010,Isenhower2010a,Zhang2010}.  A related fundamental goal is quantum manipulation of small ensembles of atoms using the blockade effect \cite{Lukin2001}.  Collective Rabi flopping was recently experimentally observed \cite{Dudin2012}.  In extended samples, {\it i.e.} atom clouds whose volume substantially exceeds the range of the Rydberg-Rydberg interactions,  a number of experiments (reviewed in Ref.~\cite{Saffman2010a}) have observed signatures of collective blockade.

In such extended samples, excitation of more than one atom per blockade volume is suppressed.  The excitation of a single atom to a Rydberg state is sufficient to shift the corresponding energy levels of its neighbors out of resonance with the exciting laser photons.  If there are $N$ atoms within the blockade volume, the effective Rabi frequency of the excitation of the blockaded ``superatom" is collectively enhanced by $\sqrt{N}$.  The Stuttgart group \cite{Heidemann2007,Heidemann2008,Low2009} has derived and experimentally confirmed the existence of non-trivial scaling relations that show how the excitation fraction depends on experimental parameters such as density and intensity.

In the main, experiments on blockade in extended samples  have been done in the absence of applied fields, so that the Rydberg-Rydberg interactions are dominated by $r^{-6}$ van der Waals interactions\cite{Walker2008}, where $r$ is the interatomic separation.  These can be tuned with DC electric fields to give F\"{o}rster resonances, with potentially dramatic increases in interaction strength.  A variety of experiments have utilized this capability, as reviewed in \cite{Comparat2010}.

Rydberg atoms {also} interact strongly with  microwave electric fields.  In the context of laser cooled Rydberg atoms, microwave spectroscopy has been used to probe energy level shifts and atomic interactions \cite{Younge2010, Anderson2011, Park2011}. Microwaves were utilized to transfer population to nearby states \cite{Maeda2011,Afrousheh2004},  and to investigate ionization of Rydberg states \cite{Park2011b}.  


Microwaves can also be used to increase the interaction strength  between Rydberg atoms.  The strong coupling of Rydberg atoms to resonant microwave fields makes it easy to dress Rydberg atoms in superpositions of opposite parity states, so that the atoms possess oscillating dipoles and thereby interact via a dipole-dipole interaction of comparable strength ($\sim n^4(ea_0)^2/r^3$) to the F\"{o}rster interaction.  The first evidence for enhanced dipole-dipole interactions from resonant microwaves was recently presented
in the context of EIT in cold Rydberg gases \cite{Tanasittikosol2011}.  In that work, Autler-Townes dressing led to enhanced suppression of EIT (as compared to van der Waals interactions)  due to the stronger,  longer range dipole-dipole interaction. 

 In this paper we use all-optical means to study  the enhanced interactions between microwave-dressed  Rydberg atoms. Using stimulated emission detection, we observe the suppression of Rydberg excitation in a magneto-optical trap.
  By monitoring the number of Rydberg atoms created by two-photon excitation while varying the density and laser intensity, a deviation from linear dependence is observed when microwave coupling is applied. The resulting suppression in the excitation of Rydberg atoms is not only qualitatively observed, but additionally quantitatively compared to theoretical predictions using  scaling law arguments and the superatom concept. The results confirm the enhancement of Rydberg-Rydberg interactions through microwave coupling.

Our experiment, described in Sec.~\ref{sec:expt}, consists of exciting 47s Rydberg states using two-photon excitation pulses.  We observe the Rydberg production by laser deexcitation to the 6p state, detecting the subsequent 6p-6s photons.  Analysis of the transients allows measurement of the excitation rates.  In Sec.~\ref{sec:scaling}, we show how the superatom concept leads to a simple model for the dependence of Rydberg production on density and intensity, for the case of a 1/$r^3$ interaction between the atoms.  We then present a calculation of the expected interaction strengths for microwave-dressed 47s Rydberg atoms, including the effects of angular momentum structure.  We then present in Sec.~\ref{sec:suppression} our results on suppression of Rydberg excitation, comparing microwave dressed and undressed atoms, and compare to the scaling theory.


\section{Experimental Setup}\label{sec:expt}

\begin{figure}
\includegraphics[width=3.5 in]{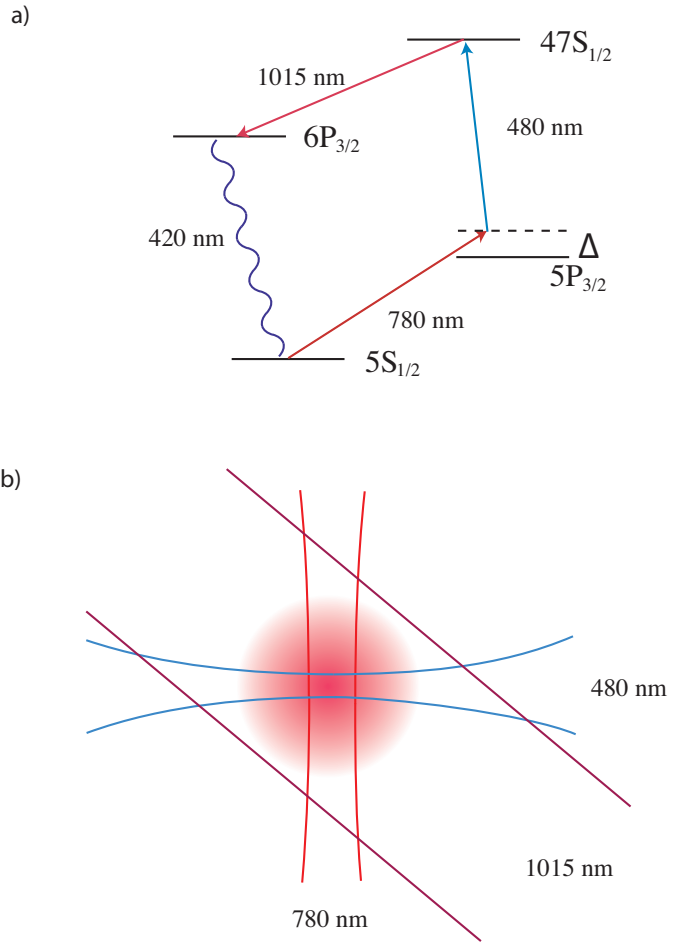}
\caption{(Color online) a) Energy levels for Rydberg atom production and detection. Two photon excitation produces Rydberg atoms, and a stimulated emission probe moves them to the 6p state where decay photons are detected. b) Experimental setup for the production of Rydberg atoms in a small spatial sample. Not to scale.}
\label{fig:elevel}
\end{figure}

The basic geometry and energy-level scheme of the experiment is shown in Fig.~\ref{fig:elevel}.  The cold atom sample is a magneto-optical trap of $^{87}$Rb atoms in the F=2 state.  In order to produce a fairly small volume of excited Rydberg atoms, we cross a 780 nm laser beam with a 50 $\mu$m waist with another 480 nm laser focused to about 13 $\mu$m.  The two lasers are tuned to produce 47s Rydberg atoms via the pathway $5s_{1/2}$(F=2)$\rightarrow5p_{3/2}\rightarrow47s$, detuned $\Delta=2\pi\times500$ MHz above the intermediate $5p_{3/2}$ F$'=3$ state.  The Rydberg atoms are de-excited by stimulated emission from a uniform intensity 1016 nm laser tuned to the $47s\rightarrow6p_{3/2}($F$'=2)$ transition, with detection of the process by observation of the subsequent $6p_{3/2}\rightarrow5s_{1/2}$ fluorescence \cite{Day2008,Brekke2008}.  

\begin{figure}
\centering
\includegraphics[scale=.6]{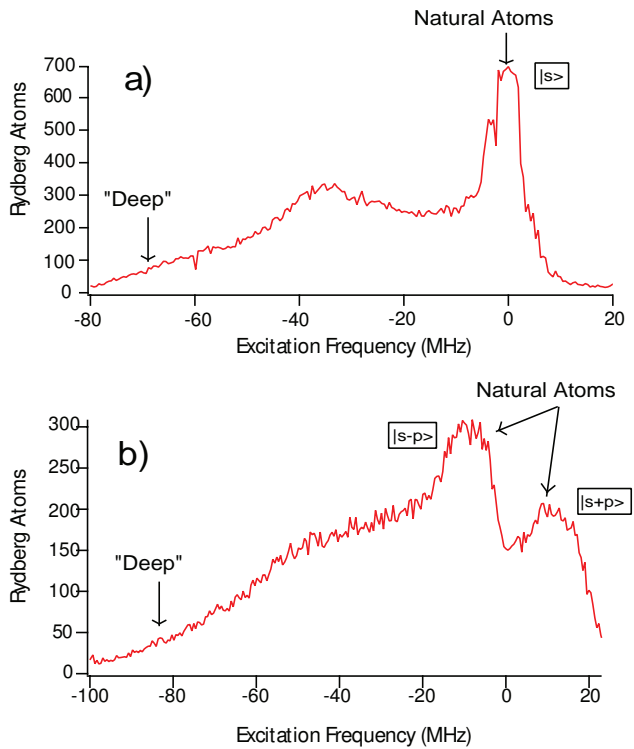}
\caption {a) Rydberg atoms are created over a large range of 480 nm frequencies due to the spatial variation of the AC Stark shift from the 780 nm beam. This could give extra spatial selection by choosing the largest shift, and hence the most intense part of the 780 nm beam, here labeled as ``Deep". b) With resonant microwaves coupling the ${47s}$ state to the ${47p_{1/2}}$ state, the resulting Autler-Townes splitting can be observed. }
\label{fig:acshift}
\end{figure}

The Rabi frequency of the 
 780 nm laser is $\Omega=2\pi \times 370$ MHz at its focus, giving rise to a spatially dependent repulsive AC Stark shift of about  ${\Omega_{780}^2}/{4 \Delta}\sim2\pi\times$ 75 MHz for the 5$s$ atoms.  This inhomogeneously broadens the two-photon excitation process, as illustrated in Fig.~\ref{fig:acshift}.  There we show the 6$p$ fluorescence as a function of the frequency of the 480 nm beam, measured relative to the free-space resonance frequency.  By tuning the 480 nm laser to the ``Deep" end of the spectrum, we selectively excite atoms at the center of the 780 beam.   We estimate that this reduces the effective waist of the 780 nm beam to about 20 $\mu$m, so that the Rydberg atoms are produced from an excitation volume of about 2$\times$20 $\mu$m$\times\pi\times$(13$\mu$m)$^2=2\times10^4$ $\mu$m$^3$.  {All results in this paper were obtained by tuning the 480 nm laser to the ``Deep" portion of the spectrum at -68 MHz.  }

Left on continuously, the repulsion of the  AC Stark shift would deplete the local atom density.  Thus  the 780 nm laser is pulsed on for only 20 $\mu$s at a time, once every 500 $\mu$s, reducing the time-averaged force by a factor of 25.  The 780 nm laser also produces significant depumping of the F=2 atoms; we estimate that about 1/4 of the atoms will be in the F=1 state at the center of the excitation region.

The 5$p_{3/2}\rightarrow47s$ Rabi frequency at the center of the crossed 480 nm beam can be calculated from the laser intensities and matrix elements, resulting in $\Omega_{480}$=2$\pi\times$11.2 MHz.  Combining this with the 780 Rabi frequency, we estimate the effective two-photon Rabi frequency to be $\Omega_{2}$=${\Omega_{780}\Omega_{480}/2\Delta}$=2$\pi\times$3.8 MHz.

As discussed in our previous work \cite{Brekke2008,Day2008}, we find that the linewidth  for excitation of Rydberg states by our lasers is about 2$\pi\times$6 MHz.  (The lasers are all external cavity diode lasers, locked to $\sim$ 10 MHz  linewidth transfer cavities that are themselves locked to Rb saturated absorption lines \cite{Brekke2009}.) The linewidth being larger than the two-photon Rabi frequency, we conclude that the excitation process is best modeled by assuming incoherent excitation.  Thus we will analyze the population dynamics for this paper in terms of populations and excitation rates, instead of using optical Bloch equations.  For excitation of a transition with a linewidth $\gamma$, the effective excitation rate is  $R = 2 \pi {\Omega^2/\gamma}$ which gives an expected peak two-photon excitation rate of $R_2=1.5 \cdot 10^7 /$s at the center of the crossed lasers.  As a result of the distribution of excitation rates within the atomic sample, the mean excitation rate is expected to be  $R_2$=$3.7 \cdot 10^6 $/s.  

{We detect Rydberg excitation by adding a third spatially uniform laser} that is tuned to be resonant on the 47$s_{1/2}\longrightarrow6p_{3/2}$ transition at 1016 nm.  {Atoms in the 6$p_{3/2}$ state decay rapidly} by spontaneous emission of 420 nm photons.  The intensity of this laser was chosen to give a stimulated emission rate of typically $R_3=3\times 10^5$/s.  This value was chosen as a compromise between small signal size (when $R_3$ is significantly smaller), or insensitivity to suppression effects (when $R_3$ is larger). 

\begin{figure}
\centering
\includegraphics[width=3.5 in]{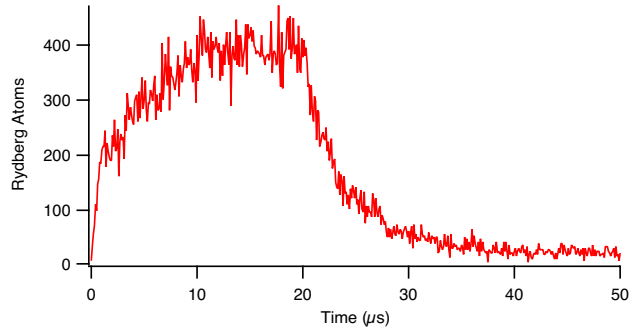}
\caption {The photon count rate is recorded vs the time after the start of the excitation pulse, and averaged for many cycles. The number of Rydberg atoms is proportional to the equilibrium count rate, and the rise/fall times of the signal are experimental measures of the excitation and stimulated emission rates. }\label{fig:Spulse}
\end{figure}

The 420 nm photons, detected by a  Hamamatsu H7360-01 photon counting module, are binned in 100 ns intervals.  The photon counter is triggered with the start of each pulse of the 780 nm beam, and the counts collected from many of these pulses averaged together. Fig. \ref{fig:Spulse} shows the result of averaging 300,000 such time sequences.  Since the decay rate from the intermediate 6$p_{3/2}$ state is significantly faster than the excitation rates used in our sample, the 420 nm photon count rate is proportional to the number of Rydberg atoms. From the count rate, we determined the number of Rydberg atoms  using the known solid angle and branching ratio ($2.4 \times 10^{-4} $), collection efficiency (12.7\%), and the  measured stimulated emission rate.

The rise and fall times of the transients in Fig.~\ref{fig:Spulse} gives an experimental measure of the excitation and stimulated emission rates through an analysis of atomic populations.  The resulting excitation is incoherent, and the optical Bloch equations can be simplified to rate equations for the three level system:

\begin{eqnarray}
{dN_g\over dt} &=& R_2 (N_r-N_g) + \Gamma_{6p} N_{6p} \nonumber\\
{dN_{r}\over dt} &=& R_2 (N_g - N_r) - R_{3} (N_{r}-N_{6p}) \nonumber \\
{dN_{6p}\over dt} &=& R_{3} (N_{r}-N_{6p}) - \Gamma_{6p} N_{6p},
\label{eqn:risemodel}
\end{eqnarray}

Since $\Gamma_{6p} >> R_3$, $N_{6p} \propto N_{r}$, allowing $N_{6p}$ to be eliminated. Solving the resulting two level system gives:
\begin{equation}
N_r= \frac{R_2}{2R_2+R_{3}}(1-e^{-(2 R_2 + R_{3})t}).
\label{eqn:nsat}
\end{equation}
The observed rise time is then equal to ${1}/(2 R_2+R_{3})$ and fall time (in the absence of excitation light) is equal to ${1}/{R_{3}}$. The stimulated emission rate  $R_3$ determined from the fit to the data agrees within a factor of four of our calculated value.  The excitation rate  $R_2$ determined from the fits is 0.6$\pm$0.2$\times 10^6$/s, smaller than the expected value by a factor of six.   


It should be noted from Fig.~\ref{fig:Spulse} that there is an initial very rapid increase in population for the first microsecond after turning on the Rydberg excitation, then a slower population increase after that.  For this data we are assuming little or no blockade, but if there is some blockade, one might explain this behavior by noting that  as excited atoms move it might be possible for previously blockaded atoms to be excited.  This effect might explain why the excitation transient in Fig.~\ref{fig:Spulse} is slower than expected.  For the following analysis, we have used the deduced values of $R_2$ and $R_3$ from data such as Fig.~\ref{fig:Spulse} at low densities where suppression effects should be small.


Finally, in order to assure an accurate correlation of photon counts to number of Rydberg atoms created, there are two concerns which must be addressed: Superradiance and collision induced ionization. We have previously seen that superradiance has played a key role in Rydberg dynamics \cite{Day2008}. However, in this experiment the small excitation size greatly reduces the number of Rydberg atoms, and we calculate the expected superradiant rates to be negligible. In this environment collision induced ionization and subsequent electron trapping present a more serious concern due to the high Rydberg densities.
Collision induced ionization and subsequent electron trapping have been shown to greatly affect the Rydberg population \cite{Li04} \cite{Walz-Flannigan2004}. The effect of ionization is often difficult to differentiate from a blockade signal \cite{Vogt2007}, and so it is important to prevent this process in our sample.  Indeed, when we switched to excitation of d-states, we found complex population dynamics on a few microsecond time scales \cite{Brekke2009}.  
S-state excitation, on the other hand, showed no sign of such effects.  This is consistent with previous observations that $s$-states ionize on much slower time scales than $d$-states do \cite{Li2005}.


In order to study how  Rydberg excitation is modified when resonant microwave fields are applied, we use an HP 83640A synthesized sweeper with a microwave horn to beam the microwaves into the vacuum chamber through a glass window. Since Rydberg atoms have very strong coupling between nearest-neighbor transitions, the states can be easily brought into the strongly coupled regime with only $\sim$0.1 mW reaching the atoms. The resulting Autler-Townes splitting can be seen in Fig. \ref{fig:acshift}b, and is well described elsewhere \cite{Autler1955}. We have examined the excitation and stimulated emission rates with microwave coupling, and seen that the microwave coupling reduces the effective rates by $\sim$2, consistent with our expectations.

The basic experiment, then, consists of studying the Rydberg population dynamics as a function of density and laser intensity, with and without microwave dressing of the Rydberg states.  To interpret our results, we use scaling arguments presented in the next section.

\section{Interaction Theory and Scaling}\label{sec:scaling}
In order to develop expectations for the scaling of Rydberg atom production in a strongly interacting gas, we use the idea of collective excitation \cite{Heidemann2007}, where the effective Rabi frequency is enhanced by the number of atoms in the blockaded region according to
\begin{equation}
\Omega_{2b} = \sqrt{N_b} \Omega_2 = \sqrt{\eta \frac{4 \pi}{3} r_b^3 } \Omega_2
\label{eqn:effexc}
\end{equation}
where $N_b$ is the number of atoms in the blockaded region, which is equal to the blockade volume $4 \pi r_b^3/3 $ times the ground state density $\eta$.
The resulting scaling in a regime where the enhanced Rabi frequency is much greater than the linewidth of the transition $\gamma$ was recently developed \cite{Low2009}. However, it is expected that in our sample the effective enhanced Rabi frequency will remain smaller than the transition linewidth. In this case, to be blockaded the atom-atom interaction energy must be comparable to or greater than the transition linewidth. For microwave coupled interactions this would give:
\begin{equation}
\frac{C_3 \beta}{r_b^3} = \gamma.
\label{eqn:uscalingw}
\end{equation}
The factor $C_3\beta$ accounts for the strength and angular dependence of the interaction, as will be discussed further below.
In the limit of strong blockade, the number of excitations contained in the extended sample volume $V$ will then be
\begin{equation}
N_s = \frac{V}{\frac{4 \pi}{3} r_b^3} = \frac{3V \gamma}{{4 \pi}C_3 \beta},
\label{eqn:sscalingw}
\end{equation}
This is independent of the density or excitation rate.

In the absence of blockade, there are potentially $N_s=\eta V$ Rydberg atoms that can be excited.  
To generalize the scaling laws for the transition between the realm where there is no blockade ($N_s \propto \eta$) and full blockade ($N_s$ constant), we use the following interpolation:
\begin{equation}
N_s=\frac{\eta V}{\sqrt{1+( \frac{3V \gamma}{{4 \pi}C_3 \beta})^2}}.
\label{eqn:sscalingg}
\end{equation}
Of course, other interpolation functions could be chosen.  This particular function was chosen for its simplicity. Several other functions with the appropriate limiting dependence were tried, and resulted in less than a 20\% change in the extracted parameters.

The relationship between the actual number of Rydberg atoms created in steady-state and the number of available excitation sites depends upon the relative rates of  excitation and stimulated emission. The steady-state number of Rydberg atoms is given by:
\begin{equation}
N_r= \frac{R_{2b}}{2 R_{2b} +R_{3}} N_s.
\label{eqn:nscalingg}
\end{equation}
In the limit of strong blockade, the effective excitation rate $R_{2b}$ is collectively enhanced by the number of atoms in a blockade volume, and so
can similarly be generalized as
\begin{equation}
R_{2b} = \sqrt{1+\left(\frac{\eta {4 \pi} C_3 \beta}{{3}\gamma}\right)^2} R_2,
\label{eqn:effrateg}
\end{equation}
so Eqn. \ref{eqn:nscalingg} can be written explicitly  as
\begin{equation}
N_r= \frac{R_{2} V \eta}{2 R_{2} \sqrt{1+\left(\frac{\eta {4 \pi} C_3 \beta}{{3}\gamma}\right)^2} +R_{3}}.
\label{eqn:nscalinggf}
\end{equation}

With this scaling relation, the number of Rydberg atoms expected can be determined for a particular set of {$\eta$, $R_2$, $R_3$, V, and $C_3 \beta$}, giving a means to quantitatively examine the number of Rydberg atoms created as a function of several experimental parameters.

It is also worthwhile to consider the scaling of the change in Rydberg atom number, ${dN_r}/{dt}$, at small times when the Rydberg state has minimal population. Following the line of thought described above,
\begin{equation}
\frac{dN_r}{dt} = N_s R_{2b},
\end{equation}
which in the most general case can be written
\begin{equation}
\frac{dN_r}{dt} = \frac{\eta V}{\sqrt{1+\left(\frac{\eta{4 \pi} C_3 \beta}{3\gamma}\right)^2}} \frac{2\pi\Omega_{2b}^2}{\sqrt{\gamma^2 +2 \Omega_{2b}^2}},
\label{eqn:dNdtscalingg}
\end{equation}
As already discussed, for the densities available in our experiment $\gamma \gg \Omega_{2b}$. This results in the simplification of Eqn. \ref{eqn:dNdtscalingg} to
\begin{equation}
\frac{dN_r}{dt} = \frac{2\pi \eta V \Omega_2^2}{\gamma},
\label{eqn:dNdtscalingw}
\end{equation}
which would scale linearly in both the ground state density and the 480 nm power. This is a different scaling than the regime where the enhanced Rabi frequency is larger than the transition linewidth, where a non-linear scaling is seen\cite{Low2009}.


We now  calculate  the atom-atom interaction strength for use in analyzing our results. When the microwaves are off, the Rydberg atoms interact via van der Waals forces, which  have been considered elsewhere \cite{Walker2008}. The interaction energy is given by
\begin{equation}
U_6(R)_{vdW}=\frac{C_6}{r^6},
\end{equation}
with relevant numbers given in Table~\ref{table:strengths}.

\begin{figure}[tb]
\includegraphics[width=3.2 in]{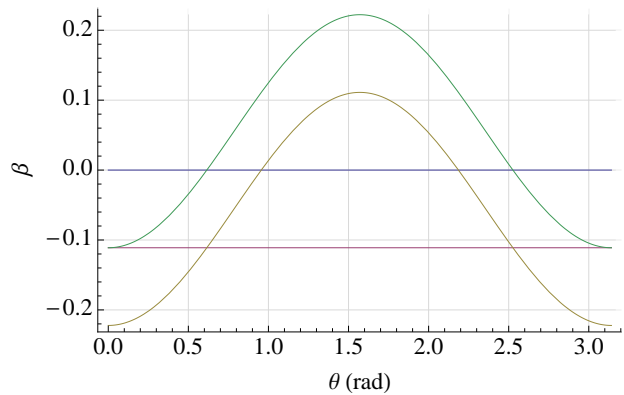}
\caption {The microwave-induced portion of the interaction energy as a function of the angle between the two dipoles for the possible angular momentum states for microwave coupling on ${s_{1/2}} \rightarrow {p_{1/2}}$. A state exists with zero microwave-induced interaction energy, which therefore experiences only weaker van der Waals interactions. The ${s_{1/2}} \rightarrow {p_{3/2}}$ coupling does not have this zero.}
\label{fig:channels}
\end{figure}

 For microwave dressed Rydberg states, the atoms acquire  oscillating dipole moments.  The time-averaged force between the dipoles is non-zero and gives an effective interaction.  Neglecting spin, dressed $s$-state  atoms would experience a dipole-dipole interaction of the classic form
\begin{equation}
U_3={C_3\over r^3}P_2(\cos\theta)
\end{equation}
where $\theta$ is the angle between the interatomic axis and the polarization of the microwave field, and $P_2$ is a Legendre polynomial.  Including the degeneracies associated with spin modifies this simple picture.  In the Appendix we consider the case of dressing on an $s_{1/2}\rightarrow p_{1/2}$ transition, as we have done in this experiment.  The results are summarized in Fig.~\ref{fig:channels}.  There are now four different interaction potentials, with corresponding eigenstates that are superpositions of the  four possible $\ket{m_1m_2}$ combinations of angular momentum states of the pair of atoms.  Two of the potentials have strong angular dependences, but the other two are isotropic, and one of them has zero dipole-dipole interaction.  In general, we express the angular averaged strength of the interaction via the parameter $\beta$ introduced in Eq. \ref{eqn:uscalingw}.

We have used the calculated interaction strengths to estimate the blockade radii and the number of atoms within a blockade sphere for various states (Table  \ref{table:strengths}).  Simulations suggest that a reasonable accounting for the blockade effects of the angular dependence of the microwave-induced interaction can made by taking $\beta=1/9$ for the three interaction curves of Fig.~\ref{fig:channels}, and assuming a van der Waals interaction for the $\beta=0$ curve.
  For our densities, we do not expect to see suppression at the $30s$ even with microwaves, while the $60s$ state would show strong suppression with either van der Waals or microwave coupled interactions.  The $47s$ level presents a compromise where the difference between the two cases is more pronounced, and hence the $47s$ state is used in the experiments described here.

\begin{table}
\centering
\begin{tabular}{c| c c c| c c c| }
\hline\hline
&  \multicolumn{3}{c|}{Van der Waals}  & \multicolumn{3}{c|}{Microwave Coupled}  \\
State  & ~~~$C_6$~~~ & ~~$r_b$~~ & ~~$N_b$~~  & ~~~$C_3\beta$~~~ & ~~$r_b$~~ & ~~$N_b$~~ \\ \hline
$30s$ &  30 & 1.3 & 0.10 &  76 & 2.3 & 0.43 \\
$47s$ &  $7.1 \times 10^3$ & 3.3 & 1.6 &  $530$ & 4.1 & 3.1 \\
$60s$ &  $1.3 \times 10^5$ & 5.3 & 6.8 & $1.4 \times 10^3$ & 5.7 & 8.5 \\
\hline
\end{tabular}
\caption{The interaction strengths $C_6$(MHz $\mu m^6$) and $C_3\beta$(MHz $\mu m^3$), blockade radius $r_b$ ($\mu$m), and blockaded number $N_b$ for various Rydberg levels accessible to our experiment. These values are shown for both van der Waals interactions and microwave coupled interactions. The transition linewidth is assumed to be 6 MHz. }
\label{table:strengths}
\end{table}



\begin{figure*} [t]
\includegraphics[scale=.56]{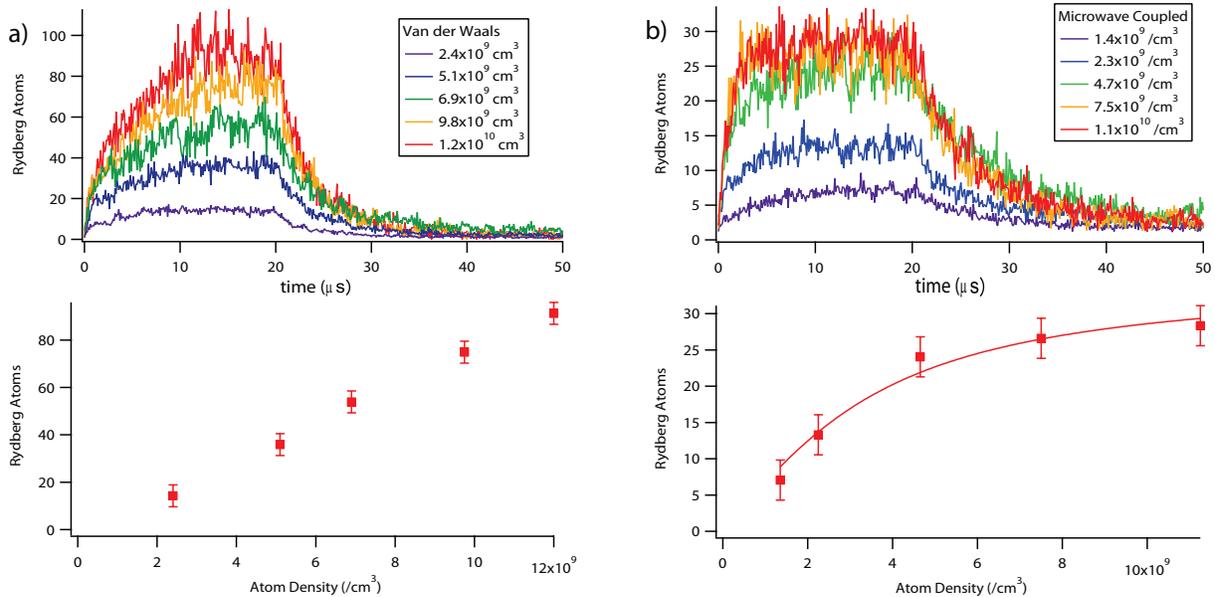}
\caption {a) An examination of the Rydberg production with only van der Waals interactions. Rydberg atom number versus ground state density shows a linear dependence, and indicates that atom-atom interactions do not limit excitation. b)  Rydberg production with microwaves coupling the ${47s}$ and ${47p_{1/2}}$ states. Rydberg atom number versus ground state density shows a clear deviation from linear dependence, and indicates that atom-atom interactions begin to suppress Rydberg atom excitation. The data is fit using Equation \ref{eqn:nscalinggf}, giving $C_3 \beta$ = 530(90) MHz $\mu$m$^3$.}
\label{fig:allNdep}
\end{figure*}

\section{Studies of Suppression}\label{sec:suppression}

We now turn to our experimental studies of suppression in the presence of resonant microwaves.

Using transients such as in Fig.~\ref{fig:Spulse}, we study Rydberg atom production as the density of atoms is changed.  We reduce the density from its maximum value by placing an iris in the MOT trapping beams, thus reducing the MOT  loading rate \cite{Brekke2009}. This technique enables variation of the number of atoms without significantly altering the volume of the atom cloud.
If no microwaves are applied, van der Waals interactions dominate. From the considerations of the previous section (Table~\ref{table:strengths}), less than 2 atoms are present in each blockaded volume at maximum density, and so minimal suppression is expected.

Figure \ref{fig:allNdep}a shows the number of Rydberg atoms versus the ground state density, with no applied microwaves.  During this series of scans, the excitation and stimulated emission rates were maintained at their maximum value:  $R_2=0.6\times10^6$/s and $R_3=0.3\times10^6$/s. 
A linear dependence is observed, supporting the estimate that atom-atom interactions do not significantly effect the excitation in this regime.

The situation changes as microwaves are applied, increasing the Rydberg-Rydberg interactions. Figure \ref{fig:allNdep}b  shows the number of Rydberg atoms versus the ground state density, while the excitation and stimulated emission rates were again  maintained at their maximum value, this time with microwaves on resonance to the $47p_{1/2}$ state. At high ground state densities, the number of Rydberg atoms produced ceases to grow linearly. This deviation from a linear dependence suggests that the increased interactions between atoms have started to suppress further Rydberg excitation as the density is increased. The slope at low density gives the effective excitation volume, while the density at which saturation begins determines the interaction strength.  
We fit the data using Eqn. \ref{eqn:nscalinggf}, and obtain an excitation volume of $2.1(4)\times 10^4$ $\mu$m$^3$ and an effective interaction strength $C_3 \beta$ of 530(90) MHz $\mu$m$^3$. Both the deduced interaction strength and the excitation volume are consistent with expectations from  Sect.~\ref{sec:expt} and Sect.~\ref{sec:scaling}. 

A prediction of the scaling model in Section~\ref{sec:scaling} is that under our conditions the slope of the Rydberg excitation curves ${dN_r}/{dt}$ should be proportional to the atom density for both the van der Waals and microwave interaction cases. Indeed, for both the van der Waals and microwave coupled cases a linear dependence on the MOT density is observed, consistent with expectations for the regime where the transition linewidth is significantly larger than the effective excitation rate. If higher densities, and hence higher effective excitation rates, are achieved in the future, it will be interesting to further examine this dependence.

\begin{figure}[tb]
\includegraphics[width=3.5 in]{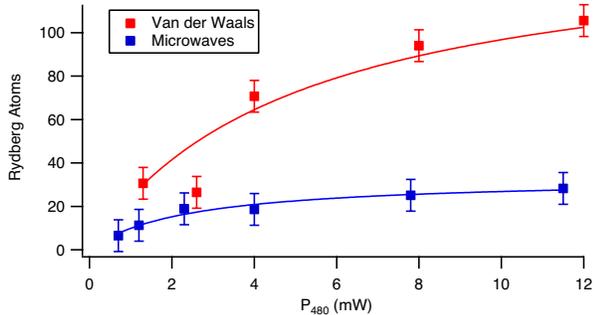}
\caption {An examination of the Rydberg atom population with microwaves coupling the ${47s}$ and ${47p_{1/2}}$ states. Rydberg atom number is plotted versus 480 nm laser power, both with and without microwaves to allow observation of the difference which increased interactions provide.  The microwave coupled data is fit using Equation \ref{eqn:nscalinggf}, giving $C_3 \beta$ = 370(100) MHz $\mu$m$^3$.}
\label{fig:microidepscans}
\end{figure}

As a second means of studying suppression of Rydberg atom excitation, we vary  the 480 nm laser power.  The data, both with and without microwave excitation,  are shown in Fig.~\ref{fig:microidepscans}.  Even in the absence of interactions, the power dependence is nonlinear due to the competition between excitation and de-excitation (see Eq.~\ref{eqn:nscalingg}).  The effect of blockade is to reduce the laser power at which the competition becomes apparent.  From  fitting the data  to Eq.~\ref{eqn:nscalinggf} we obtain an effective interaction strength $C_3 \beta$ of 370(100) MHz $\mu$m$^3$ and a volume of  $1.5(4)\times 10^4$ $\mu$m$^3$.  The deduced interaction strength is somewhat smaller (1.5 standard deviations) from our expectations and the value deduced from the density dependence data.

\section{Conclusions}

We have shown that the interactions between Rydberg atoms are increased by dressing them with resonant microwave fields, essentially giving them oscillating electric dipole moments.  In our experiment, the enhanced interactions were studied by measuring the suppression of Rydberg excitation due to the blockade effect.  We derived scaling relations valid for our experimental parameters that accounted well for the observed excitation suppression and its dependence on density and laser intensity.  The inferred interaction strengths were consistent with expectations.

Theoretically, we have shown that the Zeeman degeneracy of the dressed states alters the angular dependence of the Rydberg-Rydberg interactions.  In particular, we predict that for the $s_{1/2}\rightarrow p_{1/2}$ dressing used in our experiment, there is a particular superposition of Rydberg states that is non-interacting for all angles.  This state, which still experiences van der Waals interactions, might be more apparent in Rydberg EIT experiments, or at smaller principle quantum numbers where the van der Waals interactions are smaller relative to the microwave-dressed dipole-dipole interactions.

Interesting future experiments would be to repeat the current study using $s_{1/2}\rightarrow p_{3/2}$ dressing, where we predict that there is no non-interacting state.  In addition, our relatively broad excitation line widths of 6 MHz resulted in clearly observable but modest differences in the excitation under conditions of microwave dressing as opposed to van der Waals interactions.  With narrower excitation lines, and going to smaller principle quantum numbers, the effects should be greatly enhanced.  This suggests that it might be possible to implement asymmetric blockade schemes \cite{Saffman2009b} using time-dependent microwave fields to alternately produce strong or weak blockade.


Finally, we note that this work, as with our previous experiments and much recent work on quantum manipulation experiments using Rydberg atoms, uses photons for Rydberg detection.  It is interesting to contrast this trend with decades of experiments using ionization detection \cite{Gallagher1994}.

\begin{acknowledgments}
This research was supported by the National Science Foundation.
\end{acknowledgments}

\bibliography{MyCollection}

\appendix*

\section{Dipole-dipole interactions in dressed microwave fields}

We consider the interactions between Rydberg atoms dressed in resonant microwave fields.  For concreteness, we consider $\hat{z}$-polarized microwaves dressing an $s_{1/2}$ level (with states denoted $\ket{sm_s}$) and a $p_{1/2}$ level ($\ket{pm_p}$), but it is not difficult to extend the results to other situations.  We shall see that  the Zeeman degeneracy of the levels leads to unusual angular dependences and coupling strengths.

We assume that the microwave Rabi frequency is large compared to the dipole-dipole coupling, optical Rabi frequencies, and to Zeeman shifts.  Then, to first order, the atoms are in dressed Autler-Townes eigenstates that are linear combinations of the bare states $\ket{sm}$ and $\ket{pm}$: 
\begin{equation}
\ket{m}^\pm=\left(\ket{sm}\pm(-1)^{m-1/2}\ket{pm}e^{-i\omega t}\right)/\sqrt{2}
\end{equation}
with the $\pm$ depending on which of the two dressed states are being excited by the light.  For each atom, there are two pairs of dressed states, for $m=\pm1/2$.  Since the microwave coupling is assumed to be the strongest in the system, dipole-dipole interactions will to first order only cause transitions between the Zeeman-degenerate states within a given Autler-Townes manifold $\ket{m}^+$ or $\ket{m}^-$.

We assume that two such dressed atoms are a distance $R$ apart, with their interatomic axis oriented at an angle $\theta$ with respect to the microwave polarization.  The dipole-dipole interaction is
\begin{equation}
V={e^2\over r^3}(x'_Ax'_B+y'_Ay'_B-2z'_Az'_B)
\end{equation}
where, for example, $x'_A$ is the $x$-coordinate of the electron position operator on atom A, measured with respect to its nucleus, with the prime denoting that the operator is written in a coordinate system whose $z'$-axis is oriented along the interatomic axis. 

Considering the matrix elements of $V$ within a single dressed-state manifold, we get matrix elements like
\begin{eqnarray}
\bra{m_1 m_2}&&x'_Ax'_B\ket{m_3 m_4}=\\\nonumber
&&{1\over 4}\left(\bra{sm_1}x'_A\ket{pm_3}\bra{pm_2}x'_B\ket{sm_4}+\right.\\\nonumber
&&\left.\bra{pm_1}x'_A\ket{sm_3}\bra{sm_2}x'_B\ket{pm_4}\right)+\ldots
\end{eqnarray}
The ellipses denote terms such as  $$\bra{sm_1sm_2}x'_Ax'_B\ket{pm_3pm_4}e^{-2i\omega t}$$
that rapidly oscillate in sign and time-average to zero.

For the $s_{1/2}\rightarrow p_{1/2}$ case considered, the explicit matrix for $V$, with the basis states ordered as $\ket{1/2,1/2},\ket{1/2,-1/2},\ket{-1/2,1/2},\ket{-1/2,-1/2}$, is
\begin{equation}
V={C_3\over 9 r^3}
\left(
\begin{array}{cccc}
 -P_2 & 0 & 0 & 1-P_2 \\
 0 & -P_2 & P_2 & 0 \\
 0 & P_2 & -P_2 & 0 \\
 1-P_2 & 0 & 0 & -P_2
\end{array}
\right)\end{equation}
where $C_3=\left(e R_{n_ss}^{n_pp}\right)^2$ in the notation of Ref.~\cite{Walker2008}, and the  Legendre polynomial is  $P_{2}=(1+3\cos(2\theta))/4$.  The matrix divides into two non-interacting subspaces with $|M|=|m_A+m_B|=0$ or 1. 

The $M=0$ subspace, with all four elements proportional to $P_2$, has a zero eigenvalue; there is no energy shift from the dipole-dipole interaction for this state.  The other three eigenvalues of $V$ are non-zero and are shown in Fig.~\ref{fig:channels}.  The explicit  energies are
\begin{eqnarray}
V={C_3\over 9 r^3}\left\{0,-1,-2 P_2,1-2 P_2\right\}
\end{eqnarray}
 Note that we have neglected van der Waals interactions in this calculation, so the zero-eigenvalue state will generally still have those interactions.

We find that the case of an $s_{1/2}$ level dressed with a $p_{3/2}$ level has no zero eigenvalue case.  The energies are
\begin{eqnarray}
V&=&{C_3\over 9 r^3}\left\{\frac{3}{8} \left(2-7 P_2\right),-\frac{3}{8} \left(7 P_2-1\right)\right.\nonumber \\&&\hspace{0.25 in}\left.-\frac{3}{8} \left(P_2+1\right),-\frac{3}{8} \left(P_2+2\right)\right\}.
\end{eqnarray}
\end{document}